%% file: 2015_Summary.tex
\documentclass{IEEEtran}
\usepackage{amssymb,amsmath} 
\usepackage{graphicx}
\usepackage{url}
\usepackage[colorlinks=true, pdfstartview=FitV, linkcolor=blue, citecolor=blue, urlcolor=blue]{hyperref}
\usepackage{hyperref}

\input GKmacros.tex

\def\bib#1{\bibitem{#1}}

\begin{document}

\title{Conservation of reactive EM energy in reactive time}

\author{{Gerald Kaiser\,*\thanks{*Supported by AFOSR Grant \#FA9550-12-1-0122.}}\\
\href{http://wavelets.com}{Center for Signals and Waves}\\ Portland, OR\\
kaiser@wavelets.com
}

\maketitle

\pagestyle{empty}\thispagestyle{empty}


\begin{abstract}
The complex Poynting theorem (CPT) is extended to a canonical time-scale domain $\boldsymbol{(t,s)}$. Time-harmonic phasors are replaced by the \textit{\textbf{positive-frequency parts}} of general fields, which extend analytically to complex time $\boldsymbol{t+is}$, with $\boldsymbol{s>0}$ interpreted as a \textit{\textbf{time resolution scale.}} The real part of the extended CPT gives conservation in $\3t$ of a time-averaged field energy, and its imaginary part gives conservation in $\3s$ of a time-averaged \textit{\textbf{reactive energy.}} In both cases, the averaging windows are determined by a Cauchy kernel of width $\boldsymbol{\D t\sim \pm s}$. This \textit{\textbf{completes}}  the time-harmonic CPT, whose imaginary part is generally supposed to be vaguely `related to' reactive energy without giving a conservation law, or even an \textit{\textbf{expression,}} for the latter. The interpretation of $\3s$ as \textit{\textbf{reactive time,}} tracking the leads and lags associated with stored capacitative and inductive energy, gives a simple explanation of the \textit{\textbf{volt-ampere reactive}} (var) unit measuring reactive power: a var is simply one Joule per reactive second. The related \textit{\textbf{complex radiation impedance density}} is introduced to represent the field's local reluctance to radiate.
\end{abstract}

\section{Introduction}

The complex Poynting theorem is incomplete. While its real part is an energy conservation law, the contents of its imaginary part are less clear. Thus Jackson \ci[page 265]{J99} writes: ``It is a complex equation whose real part gives the conservation of energy for the time-averaged quantities and whose imaginary part  \textit{\textbf{relates to}}  the reactive or stored energy and its alternating flow.'' Jackson is referring to the term
\begin{align*}
i\o(\mb B\cdot\mb H^*-\mb E\cdot\mb D^*)
\end{align*}
measuring the \it balance \rm of stored magnetic and electric energies, which
determines the \it reactance \rm of a circuit or antenna under certain conditions. This suggests that the quantity in parentheses should \it somehow \rm be interpreted as a reactive energy density. But nowhere in the physics or engineering literature have I been able to find an \it exact space-time expression \rm for reactive energy in terms of the fields. I propose one here \ci{K14}.

In the process of completion, the CPT will be extended to fields with arbitrary time dependence. The \it phasors \rm which enter the time-harmonic CPT will be replaced with the \it analytic signals \rm of general fields.

\section{The CPT extended to the time-scale domain}
To minimize the notation, we work with vacuum fields and use \it natural Heaviside-Lorentz units \rm ($\e_0=\m_0=c=1$).\footnote{Where appropriate for physical clarity, $c$ will be reinstated in equations.
} 
Let $\3F\rt$ be a typical field such as $\3E\rt, \3H\rt$ or $\3J\rt$. Its \it analytic signal \rm is defined as the positive-frequency part of its Fourier representation:
\begin{align}\lab{X}
\mb F\rt=\frac1\p\int_0^\8\dd\o\, e^{i\o t}\3F_\o\orr.
\end{align}
To avoid ambiguities at $\o=0$, we assume that $\3F$ has no `DC component,' \ie $\3F_0\orr=\30$.\footnote{Static fields can be added `by hand' as in \ci{K14}.
}
If $\3F$ is real, it can be recovered fully from $\mb F$ by taking the real part,
\begin{align*}
\3F\rt=\Re\,\mb F\rt.
\end{align*}
Since the integral \eq{X} is restricted to $\o>0$, it extends analytically to the \it upper-half complex time plane \rm by
\begin{align}\lab{X1}
\mb F(\3r, t+is)=\frac1\p\int_0^\8\dd\o\, e^{i\o(t+is)}\3F_\o\orr,\qq s>0.
\end{align}
The integrand gains a \it low-pass filter \rm with transfer function $e^{-\o s}$, which suppresses frequencies $\o>>1/s$. 

A direct time-domain representation of \eq{X1} is given by the \it Cauchy transform \rm
\begin{align}\lab{X2}
\mb F(\3r,\t)=\frac i\p \ir\frac{\dd t'}{\t-t'}\,\3F(\3r, t'),\ \ \t=t+is,
\end{align}
which is a convolution with the Cauchy kernel:
\begin{align}\lab{X3}
\mb F(\3r,t+is)=C_s*\3F\rt,\ \ C_s\0t=\frac i{\p(t+is)}.
\end{align}
Roughly, $\mb F(\3r, t+is)$ depends on $\3F(\3r, t')$ in the interval $t\pm s$. Hence we say that $s$ is a \it time resolution scale \rm and call $(t,s)$ the  \it time-scale domain. \rm

To complete the CPT, note that Maxwell's equations extend analytically to complex time:
\begin{align*}
\curl\mb E(\3r,\t)&=-\pl_\t\3H(\3r,\t)\\
\curl\mb H(\3r,\t)&=\pl_\t\3E(\3r,\t)+\mb J(\3r,\t).
\end{align*}
These imply the \it extended complex Poynting theorem \rm \ci{K14}
\begin{align}\lab{CPT}
\pl_t\,\5U-i\pl_s\5X+\tfrac12\div(\mb E\times\mb H^*)=-\tfrac12\mb E\cdot\mb J^*
\end{align}
where
\begin{align*}
\5U(\3r,t,s)&=\tfrac14(|\mb H|^2+|\mb E|^2)=\text{scaled \it active \rm energy density}\\
\5X(\3r,t,s)&=\tfrac14(|\mb H|^2-|\mb E|^2)=\text{scaled \it reactive \rm energy density.}
\end{align*}

\section{Dual conservation laws}

The real and imaginary parts of \eq{CPT} are
\begin{align}
\pl_t\,\5U+\tfrac12\div\Re(\mb E\times\mb H^*)&=-\tfrac12\Re(\mb E\cdot\mb J^*)\lab{CPT1}\\
-\pl_s\5X+\tfrac12\div\Im(\mb E\times\mb H^*)&=-\tfrac12\Im(\mb E\cdot\mb J^*)\lab{CPT2}
\end{align}
The scaled densities in \eq{CPT} are defined by the convolution \eq{X3}, hence they are  \it windowed time averages \rm  over $\D t\sim \pm s$. The real part \eq{CPT1} is a conservation law in $t$ of the averaged energy at scale $s$. This is the time-domain counterpart of the \it period-averaged \rm energy density in the harmonic CPT. 

However, the imaginary part \eq{CPT2} is not a conservation law in \it time \rm but in the \it time scale \rm $s$. It states that the scaled reactive energy is conserved with respect to \it scale refinements.\rm\footnote{Due to the sign of  $-\pl_s\5X$ in \eq{CPT2} and the fact that $\5X\to0$ as $s\to\infty$, the orientation of $s$ is \it from coarse to fine scales \rm and $\5X(\3r, t, s)$ represents the \it cumulative \rm reactive energy density at all scales $s'\ge s$; see \ci[Section 5]{K14}.
}
This is rather unconventional, but we have no choice. If the CPT is extended to the \it time domain only \rm by fixing $s\=0$, then its imaginary part is \it not \rm a conservation law \ci{CBK5}. To obtain symmetry between the real and imaginary parts of the CPT, we \it must \rm extend it to $ t+is$, and then \eq{CPT} follows inevitably. Since the mathematics seems to insist that \eq{CPT} is the proper extension, let us try to make sense of it.

\section{The scale parameter as reactive time}

As a consequence of the \it non-locality \rm of \eq{X3}, $\mb F$ has a \it temporal uncertainty \rm $\D t\sim \pm s$. This creates a \it banking \rm opportunity:

\bul Given $s>0$, energy need not be conserved \it instantaneously, \rm only in an average sense over time intervals of duration $2s$. 

\bul Hence the system can \it borrow \rm energy from the future interval $\D t\sim s$ and use it to \it repay \rm energy loans made in the past interval $\D t\sim-s$, and vice-versa. Therefore \eq{CPT}  can deal with  \it energy leads and lags \rm at all time scales $s>0$. While \eq{CPT1} deals with \it average energy, \rm \eq{CPT2} deals with the \it credits and debits. \rm

\bul In the limit $s\to0$, \eq{CPT1} implies the \it real \rm Poynting theorem,\footnote{I thank Arthur Yaghjian for this observation.
}
requiring the conservation of the \it instantaneous \rm energy \ci{K14}. No such connection exists with the time-harmonic CPT since the averaging there is over a fixed period $2\p/\o>0$.

This suggests interpreting $s$ as \it reactive time, \rm measured in \it seconds reactive \rm (sr), to track energy leads and lags. The physical units in \eq{CPT2} must then be as follows:
\begin{align*}\rm
[\Im(\mb E\times\mb H^*)]=J/m^2/sr,\qq \rm [\Im(\mb E\cdot\mb J^*]=J/m^3/sr.
\end{align*}
This fully explains the \it volt ampere reactive \rm (var) unit commonly used to measure reactive power:
\begin{align*}\rm
1var=1V\cdot ar=1V\cdot C/sr=1J/sr
\end{align*}
where \it amperes reactive \rm (ar) are the units of reactive current.
Without a notion of reactive time, the var seems rather \it ad hoc. \rm

\section{Field inertia and radiation impedance}

In \ci{K11} and \ci{K12} I defined the \it inertia density \rm of an electromagnetic field by analogy with the \it mass \rm of a relativistic particle of energy $E$ and momentum $\3p$,
\begin{align}\lab{m}
m=c^{-2}\sr{E^2-c^2\3p^2}\=\sr{E^2-\3p^2}\qq (c=1).
\end{align}
For a vacuum field of energy density $U=\tfrac12(\3E^2+\3H^2)$ and momentum density $\3S=\3E\times\3H$, the inertia density is therefore
\begin{align}\lab{I}
I\rt\=\sr{U^2-\3S^2}.
\end{align}
By elementary vector algebra,
\begin{align}\lab{I1}
I\rt=\sr{R^2+X^2}
\end{align}
where
\begin{align}\lab{RX}
R\rt=|\3E\cdot\3H|\ \ \text{and}\ \  X\rt=\tfrac12(\3H^2-\3E^2).
\end{align}
I also showed that although the \it fields \rm propagate at speed $c$, their \it energy \rm flows at the local velocity
\begin{align}\lab{v}
\3v\rt=\frac{\3S\rt}{U\rt},
\end{align}
which implies that $|\3v\rt|\le c$ at every event $\rt$ and
\begin{align}\lab{v1}
|\3v\rt|=c\iff \{R\rt=X\rt=0\}.
\end{align}
The conditions $R=X=0$ thus define \it pure radiation. \rm  For a generic field, they hold only \it asymptotically \rm in the far zone; see \ci{K11}. Thus we conclude, counter-intuitively, that \it instantaneous electromagnetic energy generally flows at speeds less than $c$, even in vacuum! \rm In fact, \eq{v1} shows that the Lorentz-invariant scalars $R$ and $X$ are precisely the \it impediments \rm to radiation. Let us therefore express \eq{I1} in the suggestive form
\begin{align*}
I\rt=|Z\rt|, \ \ \text{where}\ \  Z\rt=R\rt+iX\rt
\end{align*}
may be called the \it radiation impedance density \rm by analogy with its namesake in circuit theory. Since $X$ \eq{RX} is the unscaled version of the reactive energy density $\5X$ in \eq{CPT2}, we call it the \it field reactance density. \rm This suggests that $R$ is the field analog of \it resistance \rm and should \it somehow \rm be interpreted as the \it radiation resistance density. \rm That is not as far-fetched as it may seem. Just as a particle's \it mass \rm \eq{m} impedes its \it acceleration, \rm so does a field's `complex inertia density' $Z$ impede its \it radiation. \rm  The \it scaled \rm versions of $I, Z, R, X$ are easily found to be
\begin{align}\lab{I3}
\5I(\3r, t, s)\=\sr{\5U^2-\tfrac14|\mb E\times\mb H^*|^2}=|\5Z(\3r, t, s)|
\end{align}
where
\begin{align}\lab{Z2}
\5Z=\5R+i\5X,\ \  \5R=\tfrac12|\mb E\cdot\mb H|,\ \ \5X=\tfrac14(|\mb H|^2-|\mb E|^2).
\end{align}

\end{document}

%% file: GKmacros.tex
\def\={\equiv} 
\def\div{\nabla\cdot } 
\def\pl{\partial}

\newcommand{\rt}{(\3r,t)}

\newcommand{\mb}[1]{{\mathbf #1}}

\newcommand{\bib}{\bibitem}

\newcommand{\ci}{\cite}
\newcommand{\lab}{\label}

\newcommand{\eq}{\eqref}

\newcommand{\harr}[1]{\smash{\mathop{\hbox to .5in{\ \rightarrowfill\ }}
      \limits^{#1}}}

\newcommand{\3}[1]{{\boldsymbol #1}}

\newcommand{\8}{\infty}

\def\e{\varepsilon}

\def\m{\mu} 

\def\o{\omega} 
\def\p{\pi}

\def\t{\tau}

\def\D{\Delta}

\newcommand{\bul}{$\bullet\ $}

\newcommand{\db}{{\,{\rm d}\kern-1.6ex-}}
\newcommand{\dir}{{\pl\kern-1.2ex {/}}}
\newcommand{\dd}{{\rm d}}

\newcommand{\curl}{\nabla\times}

\newcommand{\ie}{{\it i.e., }}

\def\iff{\ \Leftrightarrow\ }

\newcommand{\ir}{\int_{-\infty}^\infty}  

\newcommand{\lra}{\leftrightarrow}

\newcommand{\plra}{\pl^{\kern-1.25ex^\lra}}
\newcommand{\qq}{\quad}

\newcommand{\sr}{\sqrt}

\newcommand{\orr}{{(\3r)}}

\def\XXint#1#2#3{{\setbox0=\hbox{$#1{#2#3}{\int}$}
     \vcenter{\hbox{$#2#3$}}\kern-.5\wd0}}

\def\ED{\end{document}}